%% file: Template.tex
\documentclass{article}
\usepackage{spconf,amsmath,graphicx}
\usepackage{tikz}
\usepackage{amssymb}
\usepackage{multirow, makecell}
\usepackage{booktabs}

\usetikzlibrary{shapes}
\usetikzlibrary{automata}
\usetikzlibrary{calc,arrows,fit,positioning}
\usetikzlibrary{arrows.meta}
\usetikzlibrary{decorations.markings}
\usetikzlibrary{backgrounds}

\newlength{\nodesep}

\usepackage{environ}
\NewEnviron{FitToWidth}[1][\textwidth]{%
\begin{center}
  \makebox[#1]{\resizebox{#1}{!}{\BODY}}%
\end{center}
}

\newcommand{\F}{\mathcal{F}}
\newcommand{\yhat}{\hat{y}}
\definecolor{greenish}{rgb}{0.0, 0.5, 0.0}

\newcolumntype{R}[1]{>{\raggedleft\arraybackslash}p{#1}}

\tikzset{
  mainnode/.style = {shape=circle, draw, align=center,
  top color=white, bottom color=blue!20},
  root/.style     = {mainnode, font=\normalsize, bottom color=red!30},
  env/.style      = {mainnode, font=\ttfamily\normalsize},
  textnode/.style     = {font=\small},
  pointnode/.style = {shape=circle,draw,color=blue!70,fill,inner sep=0,minimum size=2pt},
  dummy/.style={circle,draw},
  input/.style = {shape=rectangle, draw, align=center, top color=white, bottom color=blue!20},
  hidden/.style = {shape=rectangle, draw, align=center},
  lstm/.style = {shape=rectangle, draw, align=center, top color=white, bottom color=red!20},
  output/.style = {shape=rectangle, draw, align=center, top color=white, bottom
  color=red!30},
  sup/.style = {shape=rectangle, draw, align=center, top color=white, bottom
  color=greenish!50},
  unsup/.style = {shape=rectangle, draw, align=center, top color=white, bottom
  color=red!30}
}

\title{Kaizen: Continuously improving teacher using Exponential Moving Average for semi-supervised speech recognition}
\name{\begin{tabular}{c}Vimal Manohar, Tatiana Likhomanenko, Qiantong Xu, Wei-Ning Hsu, \\ Ronan Collobert, Yatharth Saraf, Geoffrey Zweig, Abdelrahman Mohamed\end{tabular}}
\address{Facebook AI}
\begin{document}
%\ninept
%
\maketitle

\begin{abstract}
  In this paper, we introduce the Kaizen framework that uses a continuously improving teacher to generate pseudo-labels for semi-supervised speech recognition (ASR). The proposed approach uses a teacher model which is updated as the exponential moving average (EMA) of the student model parameters. We demonstrate that it is critical for EMA to be accumulated with full-precision floating point. The Kaizen framework can be seen as a continuous version of the iterative pseudo-labeling approach for semi-supervised training. It is applicable for different training criteria, and in this paper we demonstrate its effectiveness for frame-level hybrid hidden Markov model-deep neural network (HMM-DNN) systems as well as sequence-level Connectionist Temporal Classification (CTC) based models. 
  For large scale real-world unsupervised public videos in UK English and Italian languages the proposed approach i) shows more than 10\% relative word error rate (WER) reduction over standard teacher-student training; ii) using just 10 hours of supervised data and a large amount of unsupervised data closes the gap to the upper-bound supervised ASR system that uses 650h or 2700h respectively.
%   The proposed approach shows more than 10\% relative word error rate (WER) reduction over standard teacher-student training and more than 50\% relative WER reduction over a 10 hour supervised baseline, when using large scale real-world unsupervised public videos in UK English and Italian languages.
  
\end{abstract}
\begin{keywords}
 speech recognition, semi-supervised training, pseudo-labeling, low-resource, teacher-student
\end{keywords}

\section{Introduction}

Self-training~\cite{Zavaliagkos_98,thomas_13_semi,kahn2019self} is one of the most widely used approaches for semi-supervised training of automatic speech recognition (ASR). This approach uses an initial model that is called ``teacher" or ``seed" model to generate labels for unsupervised data. The generated labels are called pseudo-labels (PLs). The labeling can be done at the frame-level, which is usually in the form of soft targets or a distribution as in the case of knowledge distillation \cite{caruana_06,ba_14,hinton_15}, or at sequence-level. While there are approaches to use a distribution over sequences \cite{sequence_ts, semisup_lfmmi, hsu2020lpm} for sequence-level distillation, often only the best hypothesis sequence is used as PLs. The unsupervised data with the PLs is combined with the supervised data to train a new model. This approach is also known as ``pseudo-labeling" (PL) and serves as the baseline for semi-supervised training methods. This process can be repeated for several ``generations" to obtain better models in successive generations \cite{wessel_05}. Strong data augmentation while training the student model is shown to improve self-training and helps to avoid local optima \cite{noisy_ts,IPL,chen2020semi}. As opposed to changing the teacher model in discrete steps, i.e. after each PL generation, some recent works explored updating the model continuously and using it to generate PLs \cite{chen2020semi,IPL,slim_ipl}. In this class of approaches, we propose a new PL framework, called Kaizen. In Kaizen, we propose to use the Exponential Moving Average (EMA) of the student model as the teacher model. 

Our research is focused on semi-supervised learning for low-resource scenarios when only 1-10h of supervised data is available. 
In this paper, we make the following novel contributions: 1) We propose EMA teacher for semi-supervised ASR and empirically show that Kaizen framework in combination with data augmentation stabilizes the training even on large-scale realistic datasets with more than 10k hours of unlabeled data and only 1-10h of labeled data. 
2) We analyse the training dynamics with EMA teacher and show that for stable training it is critical for the teacher to be sufficiently far away from the student model: EMA teacher being too close to the student model causes model's collapse and divergence, while being too far leads to slow convergence. 
3) Kaizen outperforms a 10h supervised baseline and a single generation of PL by more than 50\% and 10\% relative WER reduction, respectively, with large scale real-world unsupervised public videos in UK English and Italian languages. 
4) EMA teacher combines effectively with slimIPL~\cite{slim_ipl}, an alternate approach to stabilize training, and achieves new state-of-the-art results for greedy decoding on LibriSpeech~\cite{librispeech} using labeled 10h and unlabeled 54k hours of Libri-Light~\cite{librilight} data. 

% With just 10h of labeled data and a large amount of unlabeled data, we close the gap to upper-bound ASR systems trained on 650h and 2700h, respectively, on both these languages. 
% We show that our proposed Kaizen approach in combination with data augmentation stabilizes the training even on large-scale realistic datasets with more than 10k hours of unsupervised data and only 1-10 hours of supervised data. The proposed approach shows more than 10\% relative word error rate (WER) reduction over single generation of PL and more than 50\% relative WER reduction over a 10 hour supervised baseline when using large scale real-world unsupervised public videos in UK English and Italian languages.  Using just 10h of supervised data and a large amount of unsupervised data, we close the gap to upper-bound ASR systems trained on 650h and 2700h respectively on both these languages.
% Also with Libri-Light~\cite{librilight} large scale unsupervised and 10h of supervised data Kaizen in combination with slimIPL~\cite{slim_ipl} almost closes the gap with state-of-the-art self-supervised methods on LibriSpeech~\cite{librispeech}.

In Section \ref{sec:related_works}, we compare our proposed work to related works in the literature. In Section~\ref{sec:method}, we describe the Kaizen framework and the training criteria used.
%for hybrid hidden Markov model – deep neural network (HMM-DNN) and Connectionist Temporal Classification (CTC)~\cite{ctc} modeling paradigms 
In Section~\ref{sec:experiments}, we describe the experimental setup and discuss results. In Section~\ref{sec:conclusions}, we provide our conclusions and planned future work.

\section{Related Work}
\label{sec:related_works}

EMA has been used previously for semi/self-supervised training. Temporal Ensembling~\cite{laine2016temporal_ensembling} uses EMA on network predictions while in this work we apply it on the network parameters. Mean Teacher~\cite{mean_teacher} uses EMA on parameters and consistency cost for image recognition tasks. In this work, we generalize EMA teacher to use with sequence-level loss like Connectionist Temporal Classification (CTC)~\cite{ctc} and on ASR tasks. BYOL \cite{byol} showed that EMA teacher can be used for self-supervised learning without using negative examples while our work focuses on semi-supervised learning. Multiple iterations of PL along with strong data augmentation are shown to be superior to the single generation of PL~\cite{noisy_ts, IPL}. Work~\cite{chen2020semi} extends this to continuously train a single model and use the latest model state to generate new PLs. 
In \cite{slim_ipl}, this approach was found to be unstable and prone to divergence. slimIPL algorithm~\cite{slim_ipl} gets around this via a dynamic cache containing PLs generated from the older model states. Our proposed Kaizen framework is an alternative way of stabilizing the training when using a continuously updating teacher via EMA with a sufficiently large discount factor. Independently, a concurrent work \cite{higuchi2021momentum} proposed to use EMA teacher with CTC criterion and applied it to semi-supervised ASR with 100s of hours of supervised data. In this work, we focus on the low resource scenario of 1-10hr of supervised data, and understanding the training dynamics with EMA teacher in this setting. 

\section{Method}
\label{sec:method}

\subsection{Kaizen: continuously improving teacher}

The Kaizen framework consists of a pair of models -- the teacher model and the student model -- that are trained simultaneously. The student model is trained using standard gradient-based optimization. Let its parameters 
be $\theta_t$ after $t$ updates. The teacher model parameters $\xi_t$ are updated every $\Delta$ steps as the EMA of the student model parameters:
%\footnote{More precisely, the EMA update equation is: $\xi_t = (1-\alpha) \xi_{t-\Delta} + \alpha \theta_{ \left \lfloor \frac{t}{\Delta} \right\rfloor * \Delta}$}: 
% \begin{equation}
%   \xi_t = (1-\alpha) \xi_{t-\Delta} + \alpha \theta_{ \left \lfloor \frac{t}{\Delta} \right\rfloor \Delta}, \label{eq:ema}
% \end{equation} 
\begin{equation}
  \xi_t = (1-\alpha) \xi_{t-\Delta} + \alpha \theta_t, \qquad t=n\Delta, \, n\in\mathbb{Z^+}, \label{eq:ema}
\end{equation} 
where $\alpha$ is a discount factor. A higher $\alpha$ discounts the older student models' parameters and gives more weight to the more recent student models' parameters. 
%$\Delta$ is update frequency of EMA teacher model.

In Kaizen framework, training progresses in 2 stages~-- \textit{burn-in} and continuous PL. In the \textit{burn-in} stage, the student model is trained using PLs from a previous seed model. In the continuous PL stage, the student model is trained using the PLs generated by the continuously updating teacher model. 

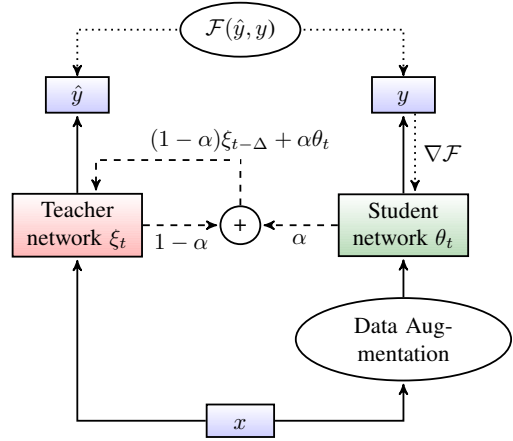
\begin{figure}[t!]
    \centering
    \begin{FitToWidth}[0.8\columnwidth] 
    \input{cpl_block_diagram}
    \end{FitToWidth}
    \vspace{-0.6cm}
    \caption{Block diagram of the Kaizen framework.}
    \vspace{-0.3cm}
    \label{fig:cpl_block_diagram}
\end{figure}

The continuous PL stage  can be described using a block diagram, Figure \ref{fig:cpl_block_diagram}. The audio features $x$ from an utterance in the unsupervised dataset is fed through both teacher and student neural network models.
For the student model, the data is augmented on-the-fly using data augmentation approaches like SpecAugment \cite{spec_augment}. The student network hidden activations are also randomly dropped using dropout \cite{srivastava2014dropout}, while dropout is not applied on the teacher network.
The resultant outputs from the teacher and student models, $\hat{y}$ and $y$ respectively,
are used to compute the loss $\mathcal{F}(\hat{y},y)$. The gradients are backpropagated 
through the student network to update its parameters $\theta_t$. 
The gradients are not backpropagated through the teacher model, 
which instead is updated as EMA of the student model parameters
with Eq.~\eqref{eq:ema}. Supervised data can be used along with unsupervised data during training.

\subsection{Exponential Moving Average (EMA)}

EMA is more commonly described by ${\lambda = 1-\alpha}$, a decay factor. 
However, we find the discount factor $\alpha$ to be more intuitive to
quantify the "distance" between the student model, which is also referred to as the online model, and its slow moving average (teacher model). The EMA model parameters
can also be unrolled as a summation over student models
after different number of updates, with each student model $\theta_i$ contributing with a weight~$w_i$ to the summation:
\begin{equation*} 
    \begin{aligned}
    \xi_t &= \sum_{i\leq t} w_i \theta_i \\
    & \triangleq \alpha \theta_t 
    + (1-\alpha)\alpha \theta_{t-\Delta} 
    + \dots 
    + {(1-\alpha)}^n \alpha \theta_{t-n\Delta} + \dots
    \end{aligned}
\end{equation*}
Another useful quantity is the half-life $\tau$ which is defined as:
% , which is defined as the number of updates behind the online model a previous student model is that contributes a weight to the teacher that is half the contribution of the online model i.e.
% i.e. 
% \begin{equation}
% \begin{aligned}
%     \tau &:= w_{t-\tau} = \frac{w_t}{2}
%      = -\Delta \frac{\ln{2}}{\ln\left(1-\alpha\right)} \label{eq:half_life}
% \end{aligned} 
% \end{equation}
\begin{equation}
\begin{aligned}
     w_{t-\tau} = \frac{w_t}{2} \qquad \text{or} \qquad \tau = -\Delta \frac{\ln{2}}{\ln\left(1-\alpha\right)} \label{eq:half_life}
\end{aligned} 
\end{equation}
% The half-lives from some values of $\alpha$ and $\Delta$ are shown in Table \ref{tab:half_lives}.
Table~\ref{tab:half_lives} shows the half-lives for some $\alpha$ and $\Delta$ values.

\begin{table}[h!]
    \centering
    \begin{tabular}{l|l|r|r}
         $\lambda$ & $\alpha$ & $\Delta$ &  $\tau$ \\
         \hline \hline 
         0.99 & 0.01 & 1 & 69 \\
         0.999 & 0.001 & 1 & 693 \\
         0.9999 & 0.0001 & 1 & 6931 \\
         0.999 & 0.001 & 10 & 6928 \\
         0.975 & 0.0025 & 10 & 2769 \\
         \hline
    \end{tabular}    
    \vspace{-0.2cm}
    \caption{Half-lives for common values of $\alpha$ and $\Delta$.}
    \label{tab:half_lives}
\end{table}

A larger $\alpha$ or equivalently a small half-life results in the teacher model being ``too close" to the student model. This can encourage the model to produce targets that are easier for the model to predict and also leads to the model ``collapse" so that the model starts to predict just silence or CTC \texttt{<blank>} token. For $\alpha=1$ and $\Delta=1$ this is consistent\footnote{\cite{slim_ipl} regularizes training via the dynamic cache while we do it via EMA.} with the observed divergence in~\cite{slim_ipl} but is in contrast to the \cite{chen2020semi} where the authors were able to train the model successfully. However, our setup is significantly different from the setup in \cite{chen2020semi} because we have only 1-10 hours of supervised data. 
% We observed empirically that the collapse can happen even when mixing in this small amount of supervised data in each epoch of training. 
% and a 1000 times larger unsupervised data whereas in \cite{chen2020semi} the unsupervised data is only 4 times as much as the supervised data. 

A smaller $\alpha$ gives smaller weight for the recent student model which results in more stable training. However, the teacher model is more static and this can lead to worse performance. We find that for better performance and stable training the half-life should be at least 1000 or higher.

In one extreme of $\alpha=0$, the teacher model is not updated at all. This is equivalent to the single-stage PL a.k.a. teacher-student training. In the other extreme of $\alpha=1$, the teacher model is replaced with the student model every $\Delta$ updates. This is equivalent to iterative PL (IPL)~\cite{IPL}. Kaizen thus provides a generalized framework for semi-supervised training that encompasses both single-stage PL as well as IPL.

\subsection{Training criteria}

The Kaizen framework can be used with different training criteria and modeling paradigms. In this paper, we investigate two modeling paradigms.

{\bf Hybrid HMM-DNN}~~~Hybrid hidden Markov model -- deep neural network (HMM-DNN) is the simplest paradigm where the neural network predicts context-dependent character (chenone) units \cite{le2019senones} at the frame-level. Here, we train the student network to minimize the Kullback-Leibler divergence~\cite{kl_divergence} between the teacher network's chenone posterior distribution $\yhat$ and student network's chenone posterior distribution $y$, $\mathcal{F}(\hat{y}, y; x) = \mathcal{D}(\hat{y} \mid\mid y)$.
% see Eq.~\eqref{eq:kl}. 
This is similar to the case of standard teacher-student training. In our work, we take the top-$k$ posteriors from the teacher network to get at least 0.99 probability mass as done in \cite{spuru2020}.
% \begin{equation}
%     \mathcal{F}(\hat{y}, y; x) = \mathcal{D}(\hat{y} \mid\mid y)\label{eq:kl}
% \end{equation}

{\bf CTC}~~~In this paradigm, the neural network is trained with the sequence-level criterion of CTC. We train the student network
% using  the CTC loss in Eq.~\eqref{eq:ctc} of
by minimizing the conditional probability of the token sequence $\yhat$ predicted by the teacher network, $\F(\yhat, y; x) = -\log p_\theta(\yhat \mid x)$.
% Eq.~\eqref{eq:ctc}. 
In this work, we use greedy decoding as in \cite{chen2020semi, slim_ipl} where the sequence $\yhat$ is obtained by de-duplicating the output label sequence
of the teacher model and removing the \texttt{<blank>} labels.
% \begin{equation}
%     \F(\yhat, y; x) = -\log p_\theta(\yhat \mid x) \label{eq:ctc}
% \end{equation}
Alternatively, a beam-search decoding can be used to obtain~$\yhat$. However, this is computationally more expensive and we did not try it in the current work. 

\subsection{Half-precision floating-point (fp16) training}
When the models are in full-precision floating point (fp32) representation, the Kaizen framework is straightforward. However, when the models are trained with half-precision floating point (fp16) for efficiency, we found that it is critical that the EMA parameters are accumulated in fp32. This results in an extra copy of EMA parameters in fp32. Without this, there is a significant degradation relative to full fp32 training, and for some parameter settings it does no better than single generation of PL. This shows that high precision is essential to capture the small changes in the EMA model.

Note that the additional fp32 copy is only used for the EMA update step. After the update step of EMA parameters, it can be cast back to fp16 so that the forward pass through the teacher network is in fp16. This allows using 1.5 times larger batch size compared to fp32 without any loss in accuracy.

\section{Experiments}
\label{sec:experiments}
\subsection{Data Preparation}

{\bf Public videos}~~~For training data, we use de-identified public videos with no personally identifiable information (PII) in UK English and Italian languages. In this paper, we simulate a low-resource scenario by limiting to a subset of 10h of supervised data, and a more extreme scenario with just 1h of supervised data in UK English. For both these languages, we use a much larger amount of unsupervised data consisting of 75k hours for UK English and 50k hours for Italian. As an upper-bound experiment, we compare with a supervised-only setting where we have 650h for UK English and 3,700h for Italian. The supervised data is augmented 3x with speed perturbation \cite{kaldi_augment}. For evaluation, we use a 23h test set for UK English and 3 test sets for Italian -- clean, noisy and extreme -- which contain 24h, 24h and 45h of data, respectively. We use a separate 14h development set for hyper-parameter tuning for both UK English and Italian. 
For UK English, we use transcripts corresponding to 650h plus an additional 13k hours of generic English video transcripts for $n$-gram language model (LM) training. For Italian, we use the transcripts from the same 3.7k hours for $n$-gram LM training.
\\
{\bf LibriSpeech}~~~We also perform experiments using Libri-Light data~\cite{librilight}: 10h labeled Libri-Light subset and 54k hours of unlabeled audio plus LibriSpeech~\cite{librispeech} itself without labels.
The standard LibriSpeech validation sets (\textit{dev-clean} and \textit{dev-other}) are used to tune all hyper-parameters, as well as to select the best models. Test sets (\textit{test-clean} and \textit{test-other}) are used only to report final word error rate (WER).
All features are normalized to have zero mean and unit variance per input sequence before feeding them into the acoustic model. 
We report not only WER without an LM, but also WER obtained by a one-pass beam-search decoder~\cite{collobert2016wav2letter} leveraging a 4-gram word-level LM~\cite{likhomanenko2019needs} and further rescoring the beam of hypothesis with a strong Transformer LM~\cite{synnaeve2019end}, following the procedure from~\cite{synnaeve2019end}. The LMs here were trained on the official LibriSpeech language modeling data~\cite{librispeech}.

% We also report WER obtained by rescoring the beam of hypothesis generated by the one-pass decoder. Rescoring is performed with a strong word-level Transformer LM, following the procedure described in~\cite{synnaeve2019end}.
% We use open-sourced word-level LMs trained on the LS LM corpus: 4-gram~\cite{likhomanenko2019needs} and Transformer~\cite{synnaeve2019end} LMs.

% \subsection{Data preparation}
%For public videos data preparation we use a hybrid flatstart lattice-free maximum mutual information (LF-MMI) \cite{lfmmi, lfbmmi_2021} trained time-delay neural network, bi-directional long short-term memory (TDNN-BLSTM) \cite{cheng2017exploration,vijay_thesis} model. 
To prepare public videos we use a time-delay neural network, bi-directional long short-term memory~\cite{cheng2017exploration,vijay_thesis} (TDNN-BLSTM) model trained using flatstart lattice-free maximum mutual information (LF-MMI) \cite{lfmmi, lfbmmi_2021} on the 10h supervised data. 
We refer to this as the alignment model and it was used to align and segment the labeled data into 10s segments for training. The unsupervised data was pre-processed using a voice activity detection model to select only speech segments of 45s maximum duration. These segments were then decoded using the alignment model to produce machine generated transcriptions which are used as 
% reference for the data preparation stage (aligning and segmentation into 10s) and later
PLs for \textit{burn-in} stage of training. The same hybrid TDNN-BLSTM LF-MMI model is trained on 10h labeled Libri-Light data and is used to generate \textit{burn-in} PLs with the 4-gram $GB \setminus LV \setminus LS$ LM~\cite{IPL} for all LibriSpeech training data.

\subsection{Model details}

The input features to all the models are 80 dimensional Mel-scale log filterbank coefficients computed every 10ms over 25ms windows. Spectral masking (both frequency and time) is applied on-the-fly using SpecAugment except for the teacher model in Kaizen framework.
The %hybrid 
TDNN-BLSTM 
%LF-MMI 
alignment model has 2~BLSTM \cite{lstm} layers with 640 hidden units in each recurrence direction and 3~TDNN layers \cite{tdnn,vijay_tdnn} with 640 hidden units interleaved between input and first BLSTM layer, and between the 2~BLSTM layers. The modeling units are context-dependent bi-character units, each modeled with a 1-state HMM topology with state-tying done using context-dependency tree built using purely the text transcripts (no alignments) and silence inserted randomly between words as done in  \cite{e2e_lfmmi, lfbmmi_2021}. 
\\
{\bf Public videos}~~~We investigate 
% the Kaizen approach with 
two modeling paradigms~-- hybrid HMM-DNN and CTC. For hybrid HMM-DNN paradigm, the modeling units are context-dependent tri-character units, each modeled with a 1-state HMM topology with state-tying done using a context-dependency tree build using statistics from the frame-level character alignments produced by the alignment model. For CTC paradigm, we use sentence-piece \cite{sp} units. For both these paradigms, we use a 80M parameter neural network with a 2 VGG layers \cite{vgg} followed by 12 Transformer blocks (768 hidden units, 8 heads) \cite{transformer} following \cite{wang2020}. Each VGG layer sub-samples by 2 in the time-axes using max-pooling \cite{yamaguchi1990neural}, resulting in the model that outputs at a rate of 25Hz (40ms time step). 
\\
{\bf LibriSpeech}~~~Here we investigate only CTC paradigm using English alphabet letters as modeling units. The neural network strictly follows~\cite{slim_ipl}: a 1-D convolution with kernel size 7 and stride 3 followed by 36 4-head Transformer blocks (768 hidden units, 4 heads), resulting in a 270M parameter model that outputs at a rate of 33.(3)Hz (30ms time step).

\subsection{Training details}

For 1h/10h of public videos data, the hybrid TDNN-BLSTM LF-MMI trained alignment model has lower WER than the 12-layer Transformer model trained using either cross-entropy (CE) or CTC losses. For example, the dev results in Table~\ref{tab:10hr_en_gb_results} for 10h supervised with CTC paradigm is significantly worse than the hybrid model. Thus the alignment model also serves as the supervised baseline. For LibriSpeech a hybrid model also outperforms a CTC model on 10h of supervised data, however the gap is small, see Table~\ref{tab:ls_10h}.

% For the semi-supervised experiments, we have a ``pre-training" stage where we train only on unsupervised data for 150k or 200k updates. 
For the semi-supervised experiments we use only unsupervised data for both \textit{burn-in} and continuous PL stages, having in total 150k/200k updates for public videos and 500k for LibriSpeech.
During the \textit{burn-in} updates (25k for public videos and 80k for LibriSpeech), we use the PLs produced by the baseline model. For public videos, EMA is accumulated only after 15k updates. This was found to not affect the performance on Librispeech; hence on Librispeech EMA was accumulated from the beginning.
%The EMA model is started to be accumulated after 15k/0 for PV/LS updates. (Vimal: Commenting it out as this is repeated in the following line.)
%There is an \textit{EMA burn-in} stage from 15k/0 updates to 25k/80k updates where the EMA model is being updated, but the student model is still being trained using PLs from the baseline model. 
After the \textit{burn-in} updates, we switch to using PLs from continuously updated teacher model (continuous PL stage). 
We follow the continuous PL stage with a fine-tuning stage where the final student model\footnote{The final EMA teacher model is only slightly better than the final student model (0.2\% absolute WER difference). Due to the small performance difference and for consistency with experiments not using EMA, we use the student model as the final model for fine-tuning.} is fine-tuned on the supervised data only.

We use the Adam~\cite{adam} and Adagrad~\cite{duchi2011adaptive} optimizer with mixed-precision~\cite{narang2017mixed} training and gradient norm clipping at 10 and 1 for public videos and LibriSpeech, respectively.
For the supervised LF-MMI baseline, we use a learning rate that rises from 1.25e-6 to 1.25e-4 in 500 updates and then reduces by a factor of 0.5 when the valid loss improvement is less than 1e-4 relative.
We use distributed data-parallel training with batch of 40min of audio distributed across 4 GPUs. 
For semi-supervised experiments with public videos, the total batch is 17.1h distributed across 64 GPUs and learning rate rises linearly for 7.5k updates to 1.5e-4 and decreases linearly to 0.
For LibriSpeech, the total batch is 0.9h distributed across 16 GPUs and learning rate rises linearly for 64k updates to 0.03 and then is decayed by 2 once valid WER reaches the plateau.
% We use DDP with batch of 17.1h for PV and 1.8h for LS of audio distributed across 64 GPUs and 32 GPUs, respectively. 
For fine-tuning stage, a learning rate rises linearly for 500 updates and decreases linearly until 10k updates and the number of GPUs is reduced to 2-4.
% DDP with batch of 40min of audio distributed across 4 GPUs.

\subsection{Results}

% \begin{figure*}[]
%     \centering
%     \includegraphics[width=\textwidth]{kaizen_wers_full}
%     \caption{WERs during pre-training stage for different EMA parameters}
%     \label{fig:kaizen_wers_full}
% \end{figure*}

\subsubsection{Public videos}

Tables \ref{tab:10hr_en_gb_results}, \ref{tab:1hr_en_gb_results} and \ref{tab:10hr_it_results} show WER results (including LM decoding) comparing standard PL, Kaizen and IPL on 10h UK English, 1h UK English and 10h Italian public videos setups. For Kaizen, we used $\alpha=0.0025, \Delta=10$. For IPL, we used $\Delta=1000$ for UK English and $\Delta=2000$ for Italian.
We also use a hybrid model trained on 10h or 1h of supervised data as the baseline. The  WER reductions (WERR) are reported relative to this baseline for all the models. We also report performance of an upper-bound model that is trained on all the supervised data that we have access to, i.e. 650h on UK English and 2.7k hours on Italian.
On UK English setups, we show WER on dev and test sets. On Italian setup, we show WER on 3 test sets -- clean, noisy and extreme. 

\begin{table}[h!]
    \centering
    \begin{tabular}{l|c|c|c|c}
        Model & Paradigm & dev & test & WERR \\
        \hline\hline
        10h sup & Hybrid & 53.9 & 51.1 & \\
        10h sup & CTC & 74.4 & - \\
        650h sup & Hybrid & 23.3 & 22.4 & 56.3 \\
        \hline
        PL & Hybrid & 30.2 & 29.8 & 41.7 \\
        Kaizen & Hybrid & 27.3 & 26.8 & 47.6 \\
        \hline
        PL & CTC & 26.2 & 25.5 & 50.2\\
        Kaizen & CTC & 23.2 & 22.7 & \bf 55.5 \\
        IPL & CTC & 23.9 & 23.4 & 54.2 \\
        \hline
    \end{tabular}
    \vspace{-0.2cm}
    \caption{WERs on 10h UK English setup with 75k hours of unsupervised data. \label{tab:10hr_en_gb_results}}
\end{table}

\begin{table}[h!]
    \centering
     \begin{tabular}{l|c|c|c|c}
        Model & Paradigm & dev & test & WERR \\
        \hline\hline
        1h sup & Hybrid & 81.1 & 79.9 & \\
        650h sup & Hybrid & 23.3 & 22.4 & 72.0 \\
        \hline
        PL & Hybrid & 64.6 & 62.3 & 22.0 \\
        Kaizen & Hybrid & 53.4 & 53.0 & 33.7 \\
        \hline
        PL & CTC & 55.3 & 54.8 & 31.4 \\
        Kaizen & CTC & 37.2 & 35.3 & \bf 55.8 \\
        IPL & CTC & 37.9 & 36.7 & 54.1 \\
        \hline
    \end{tabular}
        \vspace{-0.2cm}
    \caption{WERs on 1hr UK English setup with 75k hours of unsupervised data. \label{tab:1hr_en_gb_results}}
\end{table}

\begin{table}[h!]
    \centering
    % \begin{tabular}{l|c|c|c|c|c}
    %     Model & dev & clean & noisy & extreme & WERR \\
    %     \hline\hline
    %     10hr Supervised (CE) & 43.9 & 39.7 & 43.9 & 60.4 \\
    %     2.7khrs Supervised (CE) & 21.1 & 14.8 & 19.2 & 30.0 & 56.4\\
    %     \hline
    %     Pseudo-labeling (CTC) & 19.2 & 13.2 & 17.2 & 26.3 & 61.4\\
    %     Kaizen (CTC) & 16.6 & 11.5 & 14.6 & 21.8 & 67.2\\
    %     IPL (CTC) & 16.7 & 11.5 & 14.6 & 21.8 & 67.2\\
    %     \hline
    % \end{tabular}
    \resizebox{\linewidth}{!}{
    \begin{tabular}{l|r|r|r|c}
        Model & clean & noisy & extreme & WERR \\
        \hline\hline
        10h sup Hybrid & 39.7 & 43.9 & 60.4 \\
        %2.7khrs ub & Hybrid & 14.8 & 19.2 & 30.0 & 56.4\\
        2700h sup & 9.3 & 11.8 & 17.2 & 73.8 \\
        \hline
        PL & 13.2 & 17.2 & 26.3 & 61.4\\
        Kaizen  & 11.5 & 14.6 & 21.8 & \bf 67.2 \\
        IPL & 11.5 & 14.6 & 21.8 & \bf 67.2 \\
        \hline
    \end{tabular}
    }    \vspace{-0.2cm}
    \caption{WERs on 10hr Italian setup with 50k hours of unsupervised data. If not stated all models are CTC-based. \label{tab:10hr_it_results}}
\end{table}

We can see from the results that Kaizen outperforms PL by more than 10\% relative on all the setups for both hybrid HMM-DNN and CTC paradigms while Kaizen is similar to or slightly better than IPL. On both UK English and Italian languages, using just 10h of supervised data and a large amount of unsupervised data we close the gap to the upper-bound ASR system that uses 650h or 2.7k hours, respectively.

\subsubsection{LibriSpeech}
Table~\ref{tab:ls_10h} shows WER results comparing Kaizen and other methods. We use a hybrid model and a CTC model trained on 10h of supervised data only as the baseline. Kaizen ($\alpha=10^{-4}, \Delta=1$) demonstrates similar performance to slimIPL~\cite{slim_ipl} ($\textrm{cache}=1000$, $p=0.1$), while a combination of slimIPL and Kaizen together ($\alpha=10^{-3},\Delta=1,\textrm{cache}=1000$, $p=0.1$) achieves better performance than individual approaches. Moreover, the combination achieves a new state-of-the-art result for the greedy decoding and almost closes the gap with state-of-the-art self-supervised methods~\cite{baevski2020wav2vec,hsu2021hubert} for decoding with a language model.

\begin{table}[h!]
    \centering
    \resizebox{\linewidth}{!}{
    %{
    \begin{tabular}{l|c|R{0.07\columnwidth}|R{0.07\columnwidth}|R{0.07\columnwidth}|R{0.07\columnwidth}}
        \multirow{2}{*}{Model} & \multirow{2}{*}{LM} & \multicolumn{2}{c}{dev} & \multicolumn{2}{c}{test} \\ 
        % \cmidrule(lr){3-4} \cmidrule(lr){5-6}
        & & clean & other & clean & other \\
        \hline\hline
        \multirow{2}{*}{10h sup Hybrid} & 4-gram & 15.9 & 37.2 & 16.6 & 38.2 \\
         & $GB \setminus LV \setminus LS$ & 15.1 & 36.3 & 15.9 & 37.1 \\
        \hline
        10h sup~\cite{slim_ipl} & 4-gram & 18.8 & 39.3 & 19.6 & 39.7 \\
        %10h sup~\cite{slim_ipl} & $GB \setminus LV \setminus LS$ & 17.6 & 38.7 & 18.0 & 38.7 \\
        % 100h sup~\cite{slim_ipl} & 4.1 & 12.4 & 4.5 & 12.7 \\
        \hline
        \hline
        \multirow{2}{*}{w2v 2.0~\cite{baevski2020wav2vec}} & - & 6.3 & 9.8 & 6.3 & 10.0 \\
         & Transformer & 2.4 & 4.8 & 2.6 & 4.9 \\
        \hline
        \multirow{2}{*}{HUBERT~\cite{hsu2021hubert}} & - & 6.8 & 9.6 & 6.7 & 9.9 \\ 
         & Transformer & 2.2 & 4.3 & 2.4 & 4.6 \\ 
        \hline
        \hline
        \multirow{2}{*}{slimIPL} & - & 5.5 & 9.4 & 5.6 & 9.9 \\
         & Transformer & 2.6 & 5.4 & 3.2 & 6.1 \\
        \hline
        \multirow{2}{*}{Kaizen} & - & 5.4 & 9.5 & 5.5 & 10.1 \\
         & Transformer & 2.5 & 5.3 & 3.0 & 6.0 \\
        \hline
        \multirow{2}{*}{Kaizen+slimIPL} & - & 5.1 & 8.2 & 5.1 & 8.8 \\
         & Transformer & 2.4 & 4.9 & 2.9 & 5.5 \\
        \hline
    \end{tabular}
    }    
    \vspace{-0.2cm}
    \caption{LibriSpeech WERs for supervised baselines and different semi/self-supervised methods trained on Libri-Light, 10h labeled and 54k hours unlabeled data. If not stated all models are CTC-based.\label{tab:ls_10h}}
\end{table}

\subsubsection{Effect of EMA parameters}

In this section, we study the effect of two EMA parameters~-- the discount factor $\alpha$ and update frequency $\Delta$. We do this investigation on the UK English videos dataset
in the hybrid HMM-DNN paradigm. The stability of training depends on the distance between teacher and student models, which for Kaizen is quantified using half-life, Eq.~(\ref{eq:half_life}). 

\begin{figure}[h!]
    \centering
    \includegraphics[width=\columnwidth]{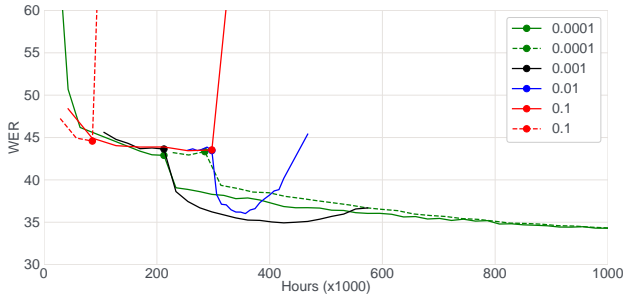}
    \vspace{-0.6cm}
    \caption{Effect of EMA discount factor $\alpha$.}
    \label{fig:kaizen_effect_of_alpha}
\end{figure}

\begin{figure}[h!]
    \centering
    \includegraphics[width=\columnwidth]{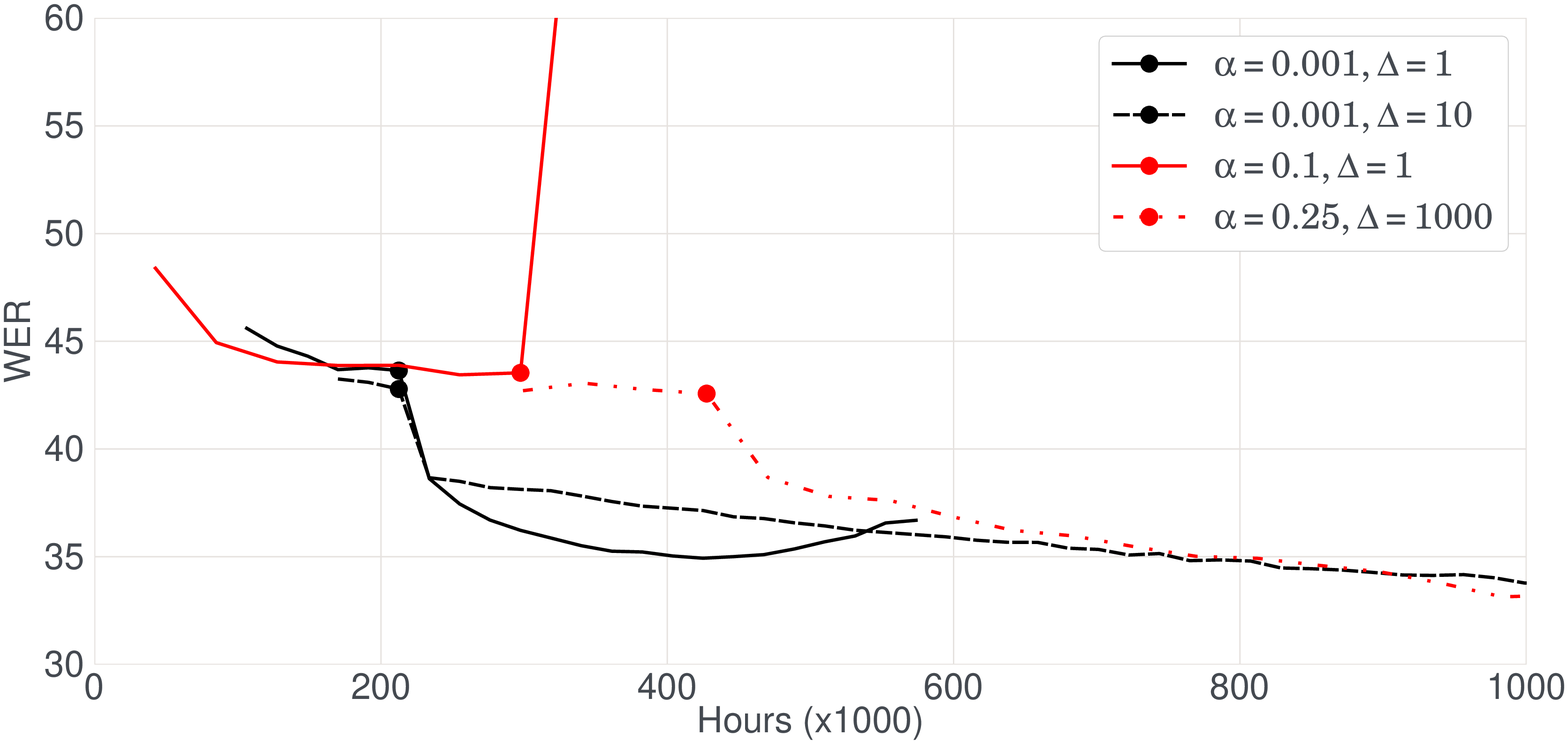}
    \vspace{-0.6cm}
    \caption{Effect of EMA update frequency $\Delta$.}
    \label{fig:kaizen_effect_of_delta}
\end{figure}

\begin{figure}[h!]
    \centering
    \includegraphics[width=\columnwidth]{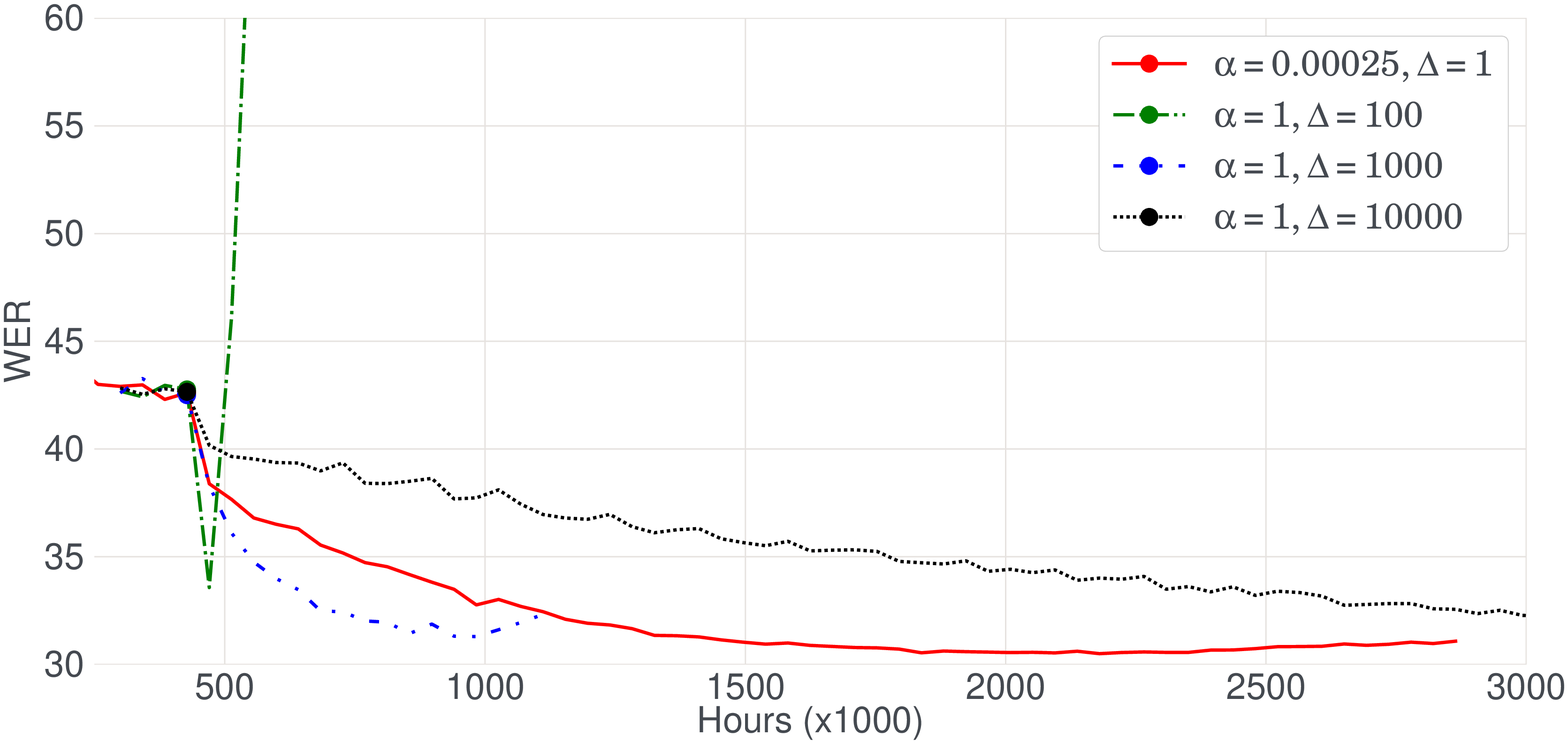}
    \vspace{-0.6cm}
    \caption{Comparing IPL and Kaizen.}
    \label{fig:kaizen_ipl}
\end{figure}

The plots in Figures \ref{fig:kaizen_effect_of_alpha}, \ref{fig:kaizen_effect_of_delta} and \ref{fig:kaizen_ipl} show the WER on UK English 10 hours supervised setup (fine-tuning stage is not included) as a function of number of training hours for various training runs. For each training run, the point where the model switches to using the continuously generated PLs is marked with a solid circle. 

\sloppy Figure \ref{fig:kaizen_effect_of_alpha} shows various training runs with $\Delta=1$ and $\alpha\in\{0.1,0.01,0.001,0.0001\}$. We see that the model diverges very quickly when $\alpha=0.1$ ($\tau=7$). The model training gets more stable progressively as we increase the $\alpha$ value towards the most stable 0.0001 ($\tau=6931$).

Figure \ref{fig:kaizen_effect_of_delta} demonstrates the effect of EMA update frequency $\Delta$. For $\alpha=0.001$, there is divergence with $\Delta=1$ ($\tau=693$), but the training is stable and WER improves continuously with $\Delta=10$ ($\tau=6928$). For higher value of $\alpha$ like 0.1 or 0.25 where half-life is less than 10 if $\Delta=1$, the training diverges almost immediately as seen for $\alpha=0.1,\Delta=1$. But even with such $\alpha$, the training is stable if $\Delta$ is increased to 1000 as seen for $\alpha=0.25,\Delta=1000$ ($\tau=2409$). We find that the post-fine-tuning model performance is also similar for similar half-life values.

Figure \ref{fig:kaizen_ipl} compares the basic Kaizen case of $\alpha=0.00025,\Delta=1$ with IPL ($\alpha=1,\Delta\in\{100,1000,10000\}$). We see that with $\Delta=100$, IPL diverges very soon after switching to using continuously generated PLs. Increasing $\Delta$ stabilizes it as seen with $\Delta=1000$ where the divergence happens only after training on 1M hours. Using a much larger $\Delta$ value of 10000 (for batch size of 17.1h and dataset of 75k hours, this is 171k hours = 2.8 epochs), the model trains stably but improves more slowly. Using typical Kaizen parameters of $\alpha=0.00025,\Delta=1$ ($\tau=2772$), the training is stable while also showing better WER after 2M hours.  

These results show that the model training is not stable unless the distance between teacher and student models is sufficiently large (half-life of more than 2000). Smaller distances i.e. smaller half-lives lead to ``collapse'' and WER degrades rapidly. In particular, we find that for updating the model continuously $\Delta=1$ as in \cite{chen2020semi} requires a small EMA discount factor to discount most recent student models. We also tried to mix-in some supervised data such that 10\% of data in each epoch is supervised. This did not help stability. We hypothesize that this is partly due to our supervised dataset being very small in the order of 1-10h. Further experiments with larger supervised datasets are needed in the future to confirm this.

\section{Conclusions and Future work}
\label{sec:conclusions}

We introduce the Kaizen framework for semi-supervised training that uses a continuously improving teacher model to generate pseudo-labels. The teacher model is updated as the exponential moving average of the student model. The proposed framework is shown as a generalization of PL and IPL. We analyzed the effect of the EMA parameters and showed that the distance between the teacher and student models is the key for effective and stable training. A small EMA half-life leads to collapse of the model and poor performance, while too large half-life leads to slow improvement. We showed that the proposed approach gives more than 10\% WERR over standard teacher-student training and performs comparatively to IPL on public videos dataset in UK English and Italian languages. We also demonstrated that Kaizen can be combined with slimIPL to achieve new state-of-the-art result for the greedy decoding and further close the gap with state-of-the-art self-supervised methods for decoding with an LM on LibriSpeech with 10h of labeled and 54k unlabeled data. %Kaizen has interesting potential for online training of ASR models without requiring additional storage of the audio.

\subsection{Future Work}

This work has explored Kaizen for Hybrid HMM-DNN and CTC based models, and we plan to explore this further for sequence-to-sequence models like RNN-T. 
%EMA sharding – EMA parameter needs to be sharded to reduce the memory consumption.
Preliminary experiments also show that scheduling EMA parameters is promising. Using larger discount factor in the beginning of training allows the teacher to forget the history and benefit from the fast improving student in the beginning. The discount factor can be later reduced to make the training more stable. While current work has shown applicability in low-resource scenario, we plan to further expand it to higher resource settings that have 100s to 1000s of hours of supervised data. The proposed approach naturally fits into online training of ASR models. 

\bibliographystyle{IEEEbib}
\bibliography{refs}

\end{document}

%% file: cpl_block_diagram.tex
\setlength{\nodesep}{0.7 cm}

\tikzset{
  mainnode/.style = {shape=circle, draw, align=center,
  top color=white, bottom color=blue!20},
  root/.style     = {mainnode, font=\normalsize, bottom color=red!30},
  env/.style      = {mainnode, font=\ttfamily\normalsize},
  textnode/.style     = {font=\small},
  pointnode/.style = {shape=circle,draw,color=blue!70,fill,inner sep=0,minimum size=2pt},
  dummy/.style={circle,draw},
  input/.style = {shape=rectangle, draw, align=center, top color=white, bottom color=blue!20},
  hidden/.style = {shape=ellipse, draw, align=center},
  teacher/.style = {shape=rectangle, draw, align=center, top color=white, bottom color=red!30},
  student/.style = {shape=rectangle, draw, align=center, top color=white, bottom color=greenish!30},
  output/.style = {shape=rectangle, draw, align=center, top color=white, bottom  color=blue!20}
}

\begin{tikzpicture}[->,>=stealth',shorten >=1pt,auto,node distance=\nodesep,
  thick]

  \node[input] at (0,0) (input) {\phantom{$x'$}$x$\phantom{$x'$}};
  \node[hidden, text width = 3\nodesep] at (2.5,1.25) (dataaug) {Data Augmentation};
  \node[student, text width = 2.5\nodesep] at (2.5,3) (student) {Student network $\theta_t$};
  \node[teacher, text width = 2.5\nodesep] at (-2.5,3) (teacher) {Teacher network $\xi_t$};
  \node[output] at (-2.5,5) (teacher_output) {\phantom{$\hat{y}$} $\hat{y}$ \phantom{$\hat{y}$}};
  \node[output] at (2.5,5) (student_output) {\phantom{$\hat{y}$} {$y$} \phantom{$\hat{y}$}};
  \node[hidden] at (0,6) (loss) {$\mathcal{F}(\hat{y},y$)};
  
  \draw (input) -| (dataaug);
  \draw (input) -| (teacher); 
  \draw (dataaug) -- (student);
  \draw (teacher) -- (teacher_output);
  \draw (student) -- (student_output);
  \draw[<-, dotted] (teacher_output) |- (loss);
  \draw[<-, dotted] (student_output) |- (loss);
  \draw[->, dotted] (student.70|-student_output.south) -- (student.70) node[midway, right] {$\nabla\mathcal{F}$};
  
  \node[hidden, ellipse] at (0,3) (ema) {+};

  \draw[->, dashed] (student) -- (ema) node[midway, below] {$\alpha$};
  \draw[->, dashed] (teacher) -- (ema) node[midway, below] {$1-\alpha$};
  
  \draw[->, dashed] (ema) |- node[pos=0.75,fill=white,inner sep=2pt]{} ++(0,1) -| (teacher.60) node[pos=0, above] {
  
  $(1-\alpha) \xi_{t-\Delta} + \alpha \theta_t$
  % { \left \lfloor \frac{t}{\Delta} \right\rfloor \Delta}
  } ;
  
\end{tikzpicture}

%% file: Template.bbl
\begin{thebibliography}{10}

\bibitem{Zavaliagkos_98}
George Zavaliagkos and Thomas Colthurst,
\newblock ``Utilizing untranscribed training data to improve performance,''
\newblock in {\em DARPA Broadcast News Transcription and Understanding
  Workshop}, 1998.

\bibitem{thomas_13_semi}
Samuel Thomas, Michael~L. Seltzer, Kenneth Church, and Hynek Hermansky,
\newblock ``Deep neural network features and semi-supervised training for low
  resource speech recognition,''
\newblock in {\em Proc. ICASSP}, 2013, pp. 6704--6708.

\bibitem{kahn2019self}
Jacob Kahn, Ann Lee, and Awni Hannun,
\newblock ``Self-training for end-to-end speech recognition,''
\newblock in {\em ICASSP}, 2020.

\bibitem{caruana_06}
Cristian Buciluundefined, Rich Caruana, and Alexandru Niculescu-Mizil,
\newblock ``Model compression,''
\newblock in {\em Proceedings of the 12th ACM SIGKDD}, 2006.

\bibitem{ba_14}
Lei~Jimmy Ba and Rich Caruana,
\newblock ``Do deep nets really need to be deep?,''
\newblock in {\em NIPS}, Cambridge, MA, USA, 2014, MIT Press.

\bibitem{hinton_15}
Geoffrey Hinton, Oriol Vinyals, and Jeffrey Dean,
\newblock ``Distilling the knowledge in a neural network,''
\newblock in {\em NIPS Deep Learning and Representation Learning Workshop},
  2015.

\bibitem{sequence_ts}
Jeremy~HM Wong and Mark Gales,
\newblock ``Sequence student-teacher training of deep neural networks,''
\newblock in {\em Interspeech}, 2016.

\bibitem{semisup_lfmmi}
Vimal Manohar, Hossein Hadian, Daniel Povey, and Sanjeev Khudanpur,
\newblock ``Semi-supervised training of acoustic models using lattice-free
  mmi,''
\newblock in {\em ICASSP}, 2018.

\bibitem{hsu2020lpm}
Wei-Ning Hsu, Ann Lee, Gabriel Synnaeve, and Awni Hannun,
\newblock ``Semi-supervised speech recognition via local prior matching,''
\newblock {\em arXiv preprint arXiv:2002.10336}, 2020.

\bibitem{wessel_05}
Frank Wessel and Hermann Ney,
\newblock ``Unsupervised training of acoustic models for large vocabulary
  continuous speech recognition,''
\newblock {\em IEEE Transactions on Speech and Audio Processing}, vol. 13, no.
  1, pp. 23--31, 2004.

\bibitem{noisy_ts}
Daniel~S Park, Yu~Zhang, Ye~Jia, Wei Han, Chung-Cheng Chiu, Bo~Li, Yonghui Wu,
  and Quoc~V Le,
\newblock ``Improved noisy student training for automatic speech recognition,''
\newblock in {\em Interspeech}, 2020.

\bibitem{IPL}
Qiantong Xu, Tatiana Likhomanenko, Jacob Kahn, Awni Hannun, Gabriel Synnaeve,
  and Ronan Collobert,
\newblock ``Iterative pseudo-labeling for speech recognition,''
\newblock in {\em Interspeech}, 2020.

\bibitem{chen2020semi}
Yang Chen, Weiran Wang, and Chao Wang,
\newblock ``{Semi-supervised ASR by End-to-end Self-training},''
\newblock in {\em Interspeech}, 2020.

\bibitem{slim_ipl}
Tatiana Likhomanenko, Qiantong Xu, Jacob Kahn, Gabriel Synnaeve, and Ronan
  Collobert,
\newblock ``slimipl: Language-model-free iterative pseudo-labeling,''
\newblock in {\em Interspeech}, 2021.

\bibitem{librispeech}
Vassil Panayotov, Guoguo Chen, Daniel Povey, and Sanjeev Khudanpur,
\newblock ``Librispeech: an asr corpus based on public domain audio books,''
\newblock in {\em ICASSP}, 2015.

\bibitem{librilight}
Jacob Kahn, Morgane Rivi{\`e}re, Weiyi Zheng, Evgeny Kharitonov, Qiantong Xu,
  et~al.,
\newblock ``Libri-light: A benchmark for asr with limited or no supervision,''
\newblock in {\em ICASSP}, 2020.

\bibitem{laine2016temporal_ensembling}
Samuli Laine and Timo Aila,
\newblock ``Temporal ensembling for semi-supervised learning,''
\newblock {\em arXiv preprint arXiv:1610.02242}, 2016.

\bibitem{mean_teacher}
Antti Tarvainen and Harri Valpola,
\newblock ``Mean teachers are better role models: Weight-averaged consistency
  targets improve semi-supervised deep learning results,''
\newblock {\em Advances in Neural Information Processing Systems}, vol. 30,
  2017.

\bibitem{ctc}
Alex Graves, Santiago Fernández, and Faustino Gomez,
\newblock ``Connectionist temporal classification: Labelling unsegmented
  sequence data with recurrent neural networks,''
\newblock in {\em ICML}, 2006.

\bibitem{byol}
Jean-Bastien Grill et~al.,
\newblock ``Bootstrap your own latent - a new approach to self-supervised
  learning,''
\newblock in {\em Advances in Neural Information Processing Systems}, 2020,
  vol.~33, pp. 21271--21284.

\bibitem{higuchi2021momentum}
Yosuke Higuchi, Niko Moritz, Jonathan~Le Roux, and Takaaki Hori,
\newblock ``{Momentum Pseudo-Labeling for Semi-Supervised Speech
  Recognition},'' 2021.

\bibitem{spec_augment}
Daniel~S Park, William Chan, Yu~Zhang, Chung-Cheng Chiu, Barret Zoph, Ekin~D
  Cubuk, and Quoc~V Le,
\newblock ``Specaugment: A simple data augmentation method for automatic speech
  recognition,''
\newblock in {\em Interspeech}, 2019.

\bibitem{srivastava2014dropout}
Nitish Srivastava, Geoffrey Hinton, Alex Krizhevsky, Ilya Sutskever, and Ruslan
  Salakhutdinov,
\newblock ``{Dropout: a simple way to prevent neural networks from
  overfitting},''
\newblock {\em The journal of machine learning research}, vol. 15, no. 1, pp.
  1929--1958, 2014.

\bibitem{le2019senones}
Duc Le, Xiaohui Zhang, Weiyi Zheng, Christian F{\"u}gen, Geoffrey Zweig, and
  Michael~L Seltzer,
\newblock ``From senones to chenones: Tied context-dependent graphemes for
  hybrid speech recognition,''
\newblock {\em ASRU}, 2019.

\bibitem{kl_divergence}
Solomon Kullback and Richard~A Leibler,
\newblock ``On information and sufficiency,''
\newblock {\em The annals of mathematical statistics}, vol. 22, no. 1, pp.
  79--86, 1951.

\bibitem{spuru2020}
Kritika Singh, Vimal Manohar, et~al.,
\newblock ``Large scale weakly and semi-supervised learning for low-resource
  video asr,''
\newblock in {\em Interspeech}, 2020.

\bibitem{kaldi_augment}
Tom Ko, Vijayaditya Peddinti, Daniel Povey, and Sanjeev Khudanpur,
\newblock ``Audio augmentation for speech recognition.,''
\newblock in {\em Interspeech}, 2015.

\bibitem{collobert2016wav2letter}
Ronan Collobert, Christian Puhrsch, and Gabriel Synnaeve,
\newblock ``Wav2letter: an end-to-end convnet-based speech recognition
  system,''
\newblock {\em arXiv preprint arXiv:1609.03193}, 2016.

\bibitem{likhomanenko2019needs}
Tatiana Likhomanenko, Gabriel Synnaeve, and Ronan Collobert,
\newblock ``Who needs words? lexicon-free speech recognition,''
\newblock {\em Proc. Interspeech 2019}, pp. 3915--3919, 2019.

\bibitem{synnaeve2019end}
Gabriel Synnaeve et~al.,
\newblock ``End-to-end asr: from supervised to semi-supervised learning with
  modern architectures,''
\newblock in {\em SAS workshop ICML}, 2020.

\bibitem{cheng2017exploration}
Gaofeng Cheng, Vijayaditya Peddinti, Daniel Povey, Vimal Manohar, Sanjeev
  Khudanpur, and Yonghong Yan,
\newblock ``An exploration of dropout with lstms.,''
\newblock in {\em Interspeech}, 2017, pp. 1586--1590.

\bibitem{vijay_thesis}
Vijayaditya Peddinti et~al.,
\newblock {\em Low latency modeling of temporal contexts for speech
  recognition},
\newblock Ph.D. thesis, Johns Hopkins University, 2017.

\bibitem{lfmmi}
Daniel Povey, Vijayaditya Peddinti, Daniel Galvez, Pegah Ghahremani, Vimal
  Manohar, Xingyu Na, Yiming Wang, and Sanjeev Khudanpur,
\newblock ``Purely sequence-trained neural networks for asr based on
  lattice-free mmi,''
\newblock in {\em Interspeech}, 2016.

\bibitem{lfbmmi_2021}
Xiaohui Zhang, Vimal Manohar, David Zhang, et~al.,
\newblock ``{On lattice-free boosted MMI training of HMM and CTC-based
  full-context ASR models},''
\newblock {\em arXiv preprint arXiv:2107.04154}, 2021.

\bibitem{lstm}
Sepp Hochreiter and J{\"u}rgen Schmidhuber,
\newblock ``Long short-term memory,''
\newblock {\em Neural computation}, vol. 9, no. 8, pp. 1735--1780, 1997.

\bibitem{tdnn}
Kevin~J Lang, Alex~H Waibel, and Geoffrey~E Hinton,
\newblock ``A time-delay neural network architecture for isolated word
  recognition,''
\newblock {\em Neural networks}, vol. 3, no. 1, pp. 23--43, 1990.

\bibitem{vijay_tdnn}
Vijayaditya Peddinti, Daniel Povey, and Sanjeev Khudanpur,
\newblock ``A time delay neural network architecture for efficient modeling of
  long temporal contexts,''
\newblock in {\em Proc. Interspeech}, 2015.

\bibitem{e2e_lfmmi}
Hossein Hadian, Hossein Sameti, Daniel Povey, and Sanjeev Khudanpur,
\newblock ``{End-to-end Speech Recognition Using Lattice-free MMI},''
\newblock in {\em Interspeech}, 2018, pp. 12--16.

\bibitem{sp}
Taku Kudo and John Richardson,
\newblock ``{S}entence{P}iece: A simple and language independent subword
  tokenizer and detokenizer for neural text processing,''
\newblock in {\em EMNLP}, 2018.

\bibitem{vgg}
Karen Simonyan and Andrew Zisserman,
\newblock ``Very deep convolutional networks for large-scale image
  recognition,''
\newblock {\em arXiv preprint arXiv:1409.1556}, 2014.

\bibitem{transformer}
Ashish Vaswani, Noam Shazeer, Niki Parmar, Jakob Uszkoreit, Llion Jones,
  Aidan~N Gomez, {\L}ukasz Kaiser, and Illia Polosukhin,
\newblock ``Attention is all you need,''
\newblock in {\em Advances in neural information processing systems}, 2017, pp.
  5998--6008.

\bibitem{wang2020}
Yongqiang Wang, Abdelrahman Mohamed, Duc Le, Chunxi Liu, Alex Xiao, et~al.,
\newblock ``Transformer-based acoustic modeling for hybrid speech
  recognition,''
\newblock in {\em ICASSP}, 2020.

\bibitem{yamaguchi1990neural}
Kouichi Yamaguchi, Kenji Sakamoto, Toshio Akabane, and Yoshiji Fujimoto,
\newblock ``A neural network for speaker-independent isolated word
  recognition,''
\newblock in {\em First International Conference on Spoken Language
  Processing}, 1990.

\bibitem{adam}
Diederik Kingma and Jimmy Ba,
\newblock ``Adam: A method for stochastic optimization,''
\newblock {\em ICLR}, 2014.

\bibitem{duchi2011adaptive}
John Duchi, Elad Hazan, and Yoram Singer,
\newblock ``Adaptive subgradient methods for online learning and stochastic
  optimization,''
\newblock {\em Journal of machine learning research}, vol. 12, no. Jul, pp.
  2121--2159, 2011.

\bibitem{narang2017mixed}
Sharan Narang, Gregory Diamos, Erich Elsen, Paulius Micikevicius, Jonah Alben,
  David Garcia, Boris Ginsburg, Michael Houston, Oleksii Kuchaiev, Ganesh
  Venkatesh, et~al.,
\newblock ``Mixed precision training,''
\newblock in {\em Int. Conf. on Learning Representation}, 2017.

\bibitem{baevski2020wav2vec}
Alexei Baevski, Yuhao Zhou, Abdelrahman Mohamed, and Michael Auli,
\newblock ``wav2vec 2.0: A framework for self-supervised learning of speech
  representations,''
\newblock {\em Advances in Neural Information Processing Systems}, vol. 33,
  2020.

\bibitem{hsu2021hubert}
Wei-Ning Hsu, Benjamin Bolte, Yao-Hung~Hubert Tsai, Kushal Lakhotia, Ruslan
  Salakhutdinov, and Abdelrahman Mohamed,
\newblock ``Hubert: Self-supervised speech representation learning by masked
  prediction of hidden units,''
\newblock {\em arXiv preprint arXiv:2106.07447}, 2021.

\end{thebibliography}
